\documentclass[12pt,preprint]{aastex}

\slugcomment{Not to appear in Nonlearned J., 45.}

\shorttitle{A constraint on the radiation efficiency...}
\shortauthors{Lu et al.}

\begin{document}


\title{On the nature of the flares from three candidate tidal disruption events}


\author{Y. Lu\altaffilmark{1}, Z. Zheng\altaffilmark{1}, S.N. Zhang\altaffilmark{2}, and Y.F. Huang\altaffilmark{3}
}
\altaffiltext{1}{National Astronomical Observatories, Chinese
 Academy of Sciences, Beijing 100012, China}
 \email{ly@bao.ac.cn}
 \altaffiltext{2}{Physics Department and
Center for Astrophysics, Tsinghua University, Beijing 100084, China}
\altaffiltext{3}{Department of Astronomy, Nanjing University,
Nanjing 210093, China}


\begin{abstract}

The X-ray flares of NGC\,5905, RX\,J1242.6-1119A, and
RX\,J1624.9+7554 observed by Chandra in 2001 and 2002 have been
suggested as the candidate tidal disruption events. The distinct
features observed from these events may be used to determine the
type of a star tidally disrupted by a massive black hole. We
investigate these three events, focusing on the differences for the
tidal disruption of a giant star and a main sequence, resulted from
their different relation between the mass and the radius. We argue
that their X-ray flare properties could be modeled by the partial
stripping of the outer layers of a solar type star. The tidal
disruption of a giant star is excluded completely. This result may
be useful for understanding the growth of a supermassive black hole
by capturing stars, versus the growth mode through continuous mass
accretion.
\end{abstract}


\section{Observation and model}
There is evidence for the existence of a supermassive black hole
with a mass of $3\times 10^6M_\odot$ at the central of our Galaxy
\citep{Sch02}. Chandra observations of NGC\,5905, RX\,J1242.6-1119A
and RX\,J1624.9+7554 in 2001 and 2002 showed that their X-ray fluxes
continued to decline at a rate consistent with the predicted
accretion rate as a function of time in the fallback phase of a
tidal disruption event by a supermassive black hole \citep{Gez03}.
\cite{Kom04} suggested that the observations are consistent with
either partial disruption of a giant (G) star, or complete
disruption of a main sequence (MS) star. Recently, by considering
the relation between the radius ($R_*$) and mass ($M_*$) of
different type of stars, Lu et al.(2006) discuss the differences
between the tidal disruption of a G and a MS star. Comparing with
the Chandra observations, we discuss in detail which type of the
disrupted star is favored for the three candidate tidal disruption
events. Throughout the paper, here we define the dimensionless
parameters $r=R_*/R_\odot$, $m=M_*/M_\odot$, and
$M_6=M_{bh}/10^6M_\odot$. We also assume that the pericenter radius
equals to the tidal radius.

Theoretically, the flare would begin when the most tightly bound
portion of the tidal debris returns to the pericenter of the star's
orbit and accretes onto the black hole. This first return would
occur at time $ t_{min} \approx 0.11 m^{-1}r^{3/2}M_6^{1/2}\,\,yr$.
The maximum return rate of debris \citep{Eva89} is
$\dot{M}_{peak}\simeq 1.4r^{-3/2}m^2M_6^{-1/2}\,M_\odot\,yr^{-1}$,
and the corresponding peak luminosity is
$L_{peak}=\epsilon\dot{M}_{peak}c^2$, where $ \epsilon \approx
5.38\times 10^{-3} r^{-1}m^{1/3}M_6^{2/3}$ is the radiation
efficiency \citep{Li02}. For a given black hole mass, one can find
that the properties of the flare, $\epsilon$ and $L_{peak}$, depend
on the properties of the disrupted star. Using the relation between
the mass and radius of the disrupted star \citep{Lu06}, we calculate
$L_{peak}$ with a supermassive black mass ranging from
$10^5$-$10^9M_\odot$. We derive $L_{peak}\sim
10^{39}$-$10^{42}\,erg\,s^{-1}$ for a disrupted giant star, and
$L_{peak}\sim 10^{43}$-$10^{45}\,erg\,s^{-1}$ for a disrupted main
sequence star, respectively.

\section{Discussion and Conclusion}
Black holes with masses of $10^6$-$10^9M_\odot$
\citep{Gez03,Fer00,Gru99} have been suggested as the sources of
tidal disruption for the flare events in NGC\,5905,
RX\,J1242.6-1119A and RX\,J1624.9+7554. Following the Chandra
observations in the 0.1-2.4keV band \citep{Hal04}, the peak
luminosity of the X-ray flares is $4\times 10^{43}\,erg\,s^{-1}$,
$1.6\times 10^{44}erg\,s^{-1}$ and $4\times 10^{43}erg\,s^{-1}$,
respectively for the above mentioned three events. Comparing with
our calculations, we find that the peak luminosity in case of a
disrupted giant star by a massive black is much lower than that
observed, but the partial disruption of a main sequence star could
satisfy the peak luminosity of the three flares. To fit the
observations, the mass fraction for a solar type star disrupted by
the black hole is 0.004 for the source of NGC\,5905, 0.016 for
RX\,J1242.6-1119A, and 0.009 for RX\,J1624.9+7554, respectively. We
summarize our conclusions as following: (1) The X-ray flares for the
three candidates of tidal disruption events are triggered by the
capture of a main sequence star; the disruption of a giant star is
excluded. (2) Our result implies that, in the regime of the black
hole growth by capturing stars, the mass growth is dominated by
capturing tidally disrupted main sequence stars. This is consistent
with the fact that the capture rate for main sequence stars is much
higher than that of giant stars.


\acknowledgements This research was supported by the National
Natural Science Foundation of China (Grants 10273011, 10573021,
10433010, and 10625313).


\end{document}